\documentclass[review]{elsarticle}

\usepackage{hyperref}
\usepackage{graphicx}
\usepackage{dcolumn}
\usepackage{bm}
\usepackage{epstopdf}
\usepackage{color}
\usepackage{textcomp}
\usepackage{amsmath}

\begin{document}
	
\title{Burnett-level discrete Boltzmann modeling of compressible flows under force}

\author[address1]{Suni Chen}
	
\author[address2,address3]{Chuandong Lin\corref{mycorrespondingauthor1}}
\cortext[mycorrespondingauthor1]{\normalsize Corresponding author: linchd3@mail.sysu.edu.cn}

\author[address1]{Demei Li}
\author[address1]{Huilin Lai\corref{mycorrespondingauthor2}}
\cortext[mycorrespondingauthor2]{\normalsize Corresponding author: hllai@fjnu.edu.cn}
	
\address[address1]{School of Mathematics and Statistics, Key Laboratory of Analytical Mathematics and Applications (Ministry of Education), Fujian Key Laboratory of Analytical Mathematics and Applications (FJKLAMA), Center for Applied Mathematics of Fujian Province (FJNU), Fujian Normal University, 350117 Fuzhou, China.}

\address[address2]{Sino-French Institute of Nuclear Engineering and Technology, Sun Yat-sen University, Zhuhai 519082, China}
	
\address[address3]{Key Laboratory for Thermal Science and Power Engineering of Ministry of Education, Department of Energy and Power Engineering, Tsinghua University, Beijing 100084, China.}

\begin{frontmatter}
\begin{abstract}
In this paper, a Burnett-level discrete Boltzmann model (DBM) is proposed for the compressible flow in a force field, and a discrete velocity set with $25$ velocities is constructed for the DBM, featuring good spatial symmetry. In the discrete Boltzmann equation, both the discrete equilibrium distribution function and the force term satisfy $25$ independent moment relations and are computed with the matrix inversion method. This approach ensures high physical accuracy, computational efficiency, and ease of implementation. Through the Chapman-Enskog expansion analysis, it is demonstrated that the current DBM can recover the Burnett equations for the compressible system under force in the continuum limit. Moreover, the DBM has the capability of capturing essential thermodynamic nonequilibrium behaviors. Finally, the model is validated through five typical benchmarks, including the free falling, Sod shock tube, sound wave, thermal Couette flow, and Rayleigh--Taylor instability.
\end{abstract}	
		
\end{frontmatter}
	
\section{Introduction}
In nature and engineering fields, fluid phenomena are ubiquitous, encompassing air, liquids, gases, and multiphase flows \cite{white2006viscous,kundu2024fluid}. These flow phenomena not only influence weather patterns, ocean currents, and blood circulation within living organisms, but also play a critical role in core processes within aerospace, aviation, energy, and chemical engineering \cite{kundu2024fluid}. To gain a deeper understanding of these complex flow mechanisms, fluid modeling and simulation research is particularly essential \cite{hirsch2007numerical}. By developing mathematical models and conducting numerical simulations, it becomes possible to predict fluid behavior, optimize system designs, and address real-world engineering challenges \cite{white2006viscous, hirsch2007numerical}.
	
The flow regimes of fluids can be classified using the Knudsen (\textit{Kn}) number, which is defined as the ratio of the molecular mean free path to the characteristic length of the system. The \textit{Kn} number reflects the degree of fluid discreteness. Based on its magnitude, fluids can be divided into four flow regimes: continuum flow, slip flow, transition flow, and free molecular flow \cite{bird1994molecular, schaaf1961flow, chapman1990mathematical}. At present, various numerical methods have been developed for these fluid flows with different \textit{Kn} numbers, and can be classified into three categories, i.e., the macroscopic, mesoscopic, and microscopic methods.

Macroscopic methods, such as those based on the Euler or Navier--Stokes (NS) equations, are widely used for their high efficiency, maturity, and broad applicability in modeling equilibrium or near-equilibrium flows. However, these methods face significant limitations in accurately capturing complex flows with pronounced nonequilibrium effects, such as rarefied gases or flows with steep gradients. Furthermore, the nonlinear nature of these methods, especially in the treatment of convective terms, requires the use of sophisticated numerical techniques. This inherent complexity limits their ability to accurately capture subtle molecular-level phenomena, which are crucial for understanding certain physical behaviors at finer scales \cite{TSIEN_1946}.
	
Microscopic methods, such as molecular dynamics, directly simulate molecular interactions and provide precise description of the microscopic properties of fluids and nonequilibrium behaviors \cite{Mao2023PECS}. These methods are particularly well-suited for investigating molecular-scale details and complex physical phenomena. However, their major limitation lies in their high computational cost, which significantly hinders their application in macroscopic-scale simulations of complex flows. Additionally, the substantial computational resources and time required by microscopic methods further constrain their feasibility for large-scale, multi-scale flow problems.
	
Mesoscopic methods serve as an effective bridge between macroscopic and microscopic approaches, providing precise descriptions of nonequilibrium flows and multi-scale interactions. Unlike macroscopic methods, the mesoscopic approaches excel in capturing significant nonequilibrium effects, while offering substantially higher computational efficiency compared to microscopic methods. Moreover, their remarkable adaptability to diverse geometries and boundary conditions makes them indispensable tools for investigating multi-phase flows, fluid instabilities, shock detonations, and other multi-scale phenomena \cite{Di_Ilio_2021}.
	
As a mesoscopic simulation approach rooted in fluid dynamics and statistical physics, the discrete Boltzmann method (DBM) has gained prominence in recent years \cite{Xu_2012, Xu_2018, Gan_2013, Lin_2016, Lin2014PRE, Zhang_2019, Gan2019FOP, Shan_2023, Chen_2022, Lai_2024, Li_2024, Lin_2014CTP,Lin_2024,Lin_2023,Lin_2019, Lai_2016PRE,Ji_2022}. By discretizing the velocity space in the Boltzmann equation, the DBM serves as a versatile tool capable of accurately describing complex fluid behaviors. The key advantages of DBM include its multi-scale simulation capability, bridging microscopic and macroscopic scales; its ability to capture nonequilibrium characteristics, enabling the analysis of thermodynamic and hydrodynamic nonequilibrium effects; its capacity to handle complex geometries and multiphase interfaces; computational efficiency suitable for large-scale parallel computing; and its broad physical applicability, extending to phenomena such as fluid instabilities, heat conduction, and chemical reactions.

In theory, it is straightforward to construct a DBM at a certain lever of physical accuracy, such as the Euler, NS, or Burnett level. The high-order DBM, compared to the Euler-level and NS-level DBMs, is better at capturing complex nonequilibrium effects, improving the prediction accuracy over a wide range of \textit{Kn} numbers. In 2018, Gan et al. proposed a two-dimensional Bhatnagar-Gross-Krook (BGK) DBM for compressible flow at the Burnett-level \cite{Gan_2018pre}. Zhang et al. introduced a novel Burnett-level discrete Boltzmann model based on the ellipsoidal statistical BGK framework to simulate nonequilibrium compressible flows \cite{ Zhang_2018}. In 2022, Zhang et al. developed a DBM based on the elliptical statistical BGK model, which considers various orders of thermodynamic nonequilibrium (TNE) effects, to study fluid flows under different levels of TNE effects \cite{Zhang_2022}. Gan et al. proposed a framework for constructing a multiscale discrete Boltzmann model for thermodynamic multiphase flows, bridging the continuum and transitional flow regimes \cite{Gan_2022}. In 2023, Wang et al. introduced a high-order DBM for multiphase flows. A force term was added to the right-hand side of the discrete Boltzmann equation to describe external forces, and another source term was introduced for molecular interactions. The equilibrium distribution function was expressed via Hermite polynomial expansion \cite{Wang_2023}. In 2024, Chen et al. proposed a multi-relaxation-time DBM for compressible non-ideal gases, investigating the impact of surface tension on Rayleigh--Taylor (RT) instability from both macroscopic and non-equilibrium perspectives \cite{Chen_2024}.
	
In this work, a two-dimensional Burnett-level DBM that incorporates a force term is proposed, making it suitable for systems under force fields. Besides, a discrete velocity model set with 25 velocities is constructed, featuring favorable spatial symmetry. This symmetry not only enhances computational efficiency and applicability but also ensures numerical accuracy and physical consistency. Moreover, this symmetry reduces computational costs while improving the stability and reliability of the simulation results. The rest of the paper is organized as follows: In Section \ref{sec:modeling}, the modeling process of the proposed DBM is introduced, including the evolution equation, the required kinetic moment relations, the matrix inversion method for calculating the equilibrium distribution function and force term, and the discrete velocity model. It is also demonstrated that the DBM can recover the Burnett equations in the continuum limit and capture various nonequilibrium effects. In Section \ref{sec:Verification and validation}, the DBM is validated numerically through five benchmarks: the free falling, Sod shock tube, sound wave, thermal Couette flow, and RT instability. Finally, Section \ref{sec:conclusion} concludes the paper.
	
\section{Discrete Boltzmann method}\label{sec:modeling}
	
\subsection{Discrete Boltzmann equation}
The Bhatnagar--Gross--Krook discrete Boltzmann equation, including the force term can be as follows\cite{Lai_2016PRE}:
	
\begin{equation}
\frac{\partial {{f}_{i}}}{\partial t}+{{\mathbf{v}}_{i}}\cdot \nabla {{f}_{i}}=-\frac{1}{\tau }\left( {{f}_{i}}-f_{i}^{\rm{eq}} \right)+{F_i}
\label{DBEquation}\text{,}
\end{equation}
where $t$ represents the time, $T$ the temperature, $\tau$ the relaxation time, $\nabla$ the nabla operator, ${\mathbf{a}}$ the acceleration, ${f}_{i}$ ($f_{i}^{\rm{eq}}$) the discrete (equilibrium) distribution function, and $i$ ($=1$, $2$, $\cdots$, $N$) the index of the discrete velocities ${\mathbf{v}}_{i}$, with $N$ being the total number of discrete velocities. On the right-hand side of the equation, ${F}_{i}$ indicates the force term, which is calculated by Eq. (\ref{Expression_force_i}).
	
The Maxwellian equilibrium distribution function ${{f}^{\rm{eq}}}$ is given by\cite{Lin2017PRE}:
\begin{equation}
		{{f}^{\rm{eq}}}=n{{\left( \frac{m}{2\pi T} \right)}^{D/2}}{{\left( \frac{m}{2\pi IT} \right)}^{1/2}}\exp \left[ -\frac{m{{\left| \mathbf{v}-\mathbf{u} \right|}^{2}}}{2T}-\frac{m{{\eta }^{2}}}{2IT} \right]
		\text{,}
		\label{Expression_feq}
\end{equation}
where $m = 1$ the molar mass, ${\mathbf{u}}$ the flow velocity, $n$ denotes the molar concentration,  ${\mathbf{v}}$ represents the particle velocity, the symbols $\eta$ is used to describe the extra internal energies corresponding to molecular vibration and/or rotation, $D = 2$ represents the spatial dimension, and $I$ stands for the extra degrees of freedom.
	
To construct a Burnett-level DBM, the following kinetic moment relations for the discrete equilibrium distribution function should be satisfied:
\begin{equation}
		\sum\nolimits_{i}{f_{i}^{\rm{eq}}}=\iint{{{f}^{\rm{eq}}}d\mathbf{v}d\eta }
		\label{Moment_feq0}
		\text{,}
\end{equation}
\begin{equation}
		\sum\nolimits_{i}{f_{i}^{\rm{eq}}{{v}_{i\alpha }}}=\iint{{{f}^{\rm{eq}}}{{v}_{\alpha }}d\mathbf{v}d\eta }
		\label{Moment_feq1}
		\text{,}
\end{equation}
\begin{equation}
		\sum\nolimits_{i}{f_{i}^{\rm{eq}}\left( v_{i}^{2}+\eta _{i}^{2} \right)}=\iint{{{f}^{\rm{eq}}}\left( {{v}^{2}}+{{\eta }^{2}} \right)d\mathbf{v}d\eta }
		\label{Moment_feq2,0}
		\text{,}
\end{equation}
\begin{equation}
		\sum\nolimits_{i}{f_{i}^{\rm{eq}}{{v}_{i\alpha }}{{v}_{i\beta }}}=\iint{{{f}^{\rm{eq}}}{{v}_{\alpha }}{{v}_{\beta }}d\mathbf{v}d\eta }
		\label{Moment_feq2}
		\text{,}
\end{equation}
\begin{equation}
		\sum\nolimits_{i}{f_{i}^{\rm{eq}}\left( v_{i}^{2}+\eta _{i}^{2} \right)v_{i\alpha }}=\iint{{{f}^{\rm{eq}}}\left( {{v}^{2}}+{{\eta }^{2}} \right){{v}_{\alpha }}d\mathbf{v}d\eta }
		\label{Moment_feq3,1}
		\text{,}
\end{equation}
\begin{equation}
		\sum\nolimits_{i}{f_{i}^{\rm{eq}}{{v}_{i\alpha }}{{v}_{i\beta }}{{v}_{i\chi }}}=\iint{{{f}^{\rm{eq}}}{{v}_{\alpha }}{{v}_{\beta }}{{v}_{\chi }}d\mathbf{v}d\eta }
		\label{Moment_feq3}
		\text{,}
\end{equation}
\begin{equation}
		\sum\nolimits_{i}{f_{i}^{\rm{eq}}\left( v_{i}^{2}+\eta _{i}^{2} \right){{v}_{i\alpha }}{{v}_{i\beta }}}=\iint{{{f}^{\rm{eq}}}\left( {{v}^{2}}+{{\eta }^{2}} \right){{v}_{\alpha }}{{v}_{\beta }}d\mathbf{v}d\eta }
		\label{Moment_feq4,2}
		\text{,}
\end{equation}
\begin{equation}
		\sum\nolimits_{i}{f_{i}^{\rm{eq}}{{v}_{i\alpha }}{{v}_{i\beta }}{{v}_{i\chi }}{{v}_{i\gamma }}}=\iint{{{f}^{\rm{eq}}}{{v}_{\alpha }}{{v}_{\beta }}{{v}_{\chi }}{{v}_{\gamma }}d\mathbf{v}d\eta }
		\label{Moment_feq4}
		\text{,}
\end{equation}
\begin{equation}
		\sum\nolimits_{i}{f_{i}^{\rm{eq}}\left( v_{i}^{2}+\eta _{i}^{2} \right){{v}_{i\alpha }}{{v}_{i\beta }}{{v}_{\chi }}}=\iint{{{f}^{\rm{eq}}}\left( {{v}^{2}}+{{\eta }^{2}} \right){{v}_{\alpha }}{{v}_{\beta }}{{v}_{\chi }}d\mathbf{v}d\eta }
		\label{Moment_feq5,3}
		\text{,}
\end{equation}
where $\eta_{i}$ is the free parameter,   $v=|\mathbf{v}|$  (${v_{i}}=|\mathbf{v}_{i}|$) represents the magnitude of the particle (discrete) velocity, and ${v}_{\alpha}$ (${v}_{i\alpha}$) is the component of the particle (discrete) velocity $\mathbf{v}$ (${{\mathbf{v}}_{i}}$) in the $\alpha$-direction. For the two dimensional system, $\alpha$, $\beta$, $\chi$, $\gamma$ = $x$ or $y$. The above nine sets of kinetic moment relations can be expressed in matrix form:
\begin{equation}
		{\mathbf{C}} \cdot {\mathbf{{f}^{\rm{eq}}}} = {\mathbf{\hat f^{\rm{eq}}}}
		\label{Matrix_form_feq}
		\text{,}
\end{equation}
where ${\mathbf{{f}^{\rm{eq}}}}=[f_{1}^{\rm{eq}},f_{2}^{\rm{eq}},\dots,f_{N}^{\rm{eq}}]^{\rm{T}}$ represents the column matrix of discrete equilibrium distribution functions, ${\mathbf{{\hat f}^{\rm{eq}}}}=[\hat f_{1}^{\rm{eq}}, \hat f_{2}^{\rm{eq}}, \dots, \hat f_{N}^{\rm{eq}}]^{\rm{T}}$ is the column matrix of equilibrium kinetic moments, and ${\mathbf{C}}$ is a square matrix whose elements depends on the discrete parameters ${\mathbf{v}}_{i}$ and ${\eta}_{i}$.
When ${\mathbf{C}}$ is invertible, Eq. (\ref{Matrix_form_feq}) can be expressed as:
\begin{equation}
		{\mathbf{{f}^{\rm{eq}}}} = {\mathbf{C}}^{-1} \cdot {\mathbf{{{\hat f}}^{\rm{eq}}}}
		\label{Expression_feq_i}
		\text{,}
\end{equation}
where ${\mathbf{C}}^{-1}$ is the inverse matrix of ${\mathbf{C}}$. Note that Eq. (\ref{Expression_feq_i}) provides the way how the discrete equilibrium distribution function is calculated.
	
Similarly, the force term ${F}_{i}$ should also satisfy the following nine sets of kinetic moment relations:
\begin{equation}
		\sum\nolimits_{i}{F_{i}}=\iint{{F}d\mathbf{v}d\eta }
		\label{Moment_force0}
		\text{,}
\end{equation}
\begin{equation}
		\sum\nolimits_{i}{F_{i}}{{v}_{i\alpha }}=\iint{{F}{{v}_{\alpha }}d\mathbf{v}d\eta }
		\label{Moment_force1}
		\text{,}
\end{equation}
\begin{equation}
		\sum\nolimits_{i}{F_{i}\left( v_{i}^{2}+\eta _{i}^{2} \right)}=\iint{{F}\left( {{v}^{2}}+{{\eta }^{2}} \right)d\mathbf{v}d\eta }
		\label{Moment_force2,0}
		\text{,}
\end{equation}
\begin{equation}
		\sum\nolimits_{i}{F_{i}{{v}_{i\alpha }}{{v}_{i\beta }}}=\iint{{F}{{v}_{\alpha }}{{v}_{\beta }}d\mathbf{v}d\eta }
		\label{Moment_force2}
		\text{,}
\end{equation}
\begin{equation}
		\sum\nolimits_{i}{F_{i}}\left( v_{i}^{2}+\eta _{i}^{2} \right)v_{i\alpha }=\iint{F\left( {{v}^{2}}+{{\eta }^{2}} \right){{v}_{\alpha }}d\mathbf{v}d\eta }
		\label{Moment_force3,1}
		\text{,}
\end{equation}
\begin{equation}
		\sum\nolimits_{i}{F_{i}{{v}_{i\alpha }}{{v}_{i\beta }}{{v}_{i\chi }}}=\iint{{F}{{v}_{\alpha }}{{v}_{\beta }}{{v}_{\chi }}d\mathbf{v}d\eta }
		\label{Moment_force3}
		\text{,}
\end{equation}
\begin{equation}
		\sum\nolimits_{i}{F_{i}}\left( v_{i}^{2}+\eta _{i}^{2} \right)v_{i\alpha }v_{i\beta }=\iint{F\left( {{v}^{2}}+{{\eta }^{2}} \right){{v}_{\alpha }}{{v}_{\beta }}d\mathbf{v}d\eta }
		\label{Moment_force4,2}
		\text{,}
\end{equation}
\begin{equation}
		\sum\nolimits_{i}{F_{i}{{v}_{i\alpha }}{{v}_{i\beta }}{{v}_{i\chi }}{{v}_{i\gamma }}}=\iint{{F}{{v}_{\alpha }}{{v}_{\beta }}{{v}_{\chi }}{{v}_{\gamma }}d\mathbf{v}d\eta }
		\label{Moment_force4}
		\text{,}
\end{equation}
\begin{equation}
		\sum\nolimits_{i}{F_{i}}\left( v_{i}^{2}+\eta _{i}^{2} \right)v_{i\alpha }v_{i\beta }v_{i\chi }=\iint{F\left( {{v}^{2}}+{{\eta }^{2}} \right){{v}_{\alpha }}{{v}_{\beta }}{{v}_{\chi }}d\mathbf{v}d\eta }
		\label{Moment_force5,3}
		\text{,}
\end{equation}
where
\begin{equation}
		{F}=-\mathbf{a}\cdot\frac{\partial {f}}{\partial \mathbf{v}}\approx -\mathbf{a}\cdot\frac{\partial {f}^{\rm{eq}}}{\partial \mathbf{v}} =-\frac{m\mathbf{a}\cdot(\mathbf{v}-\mathbf{u})}Tf^{\rm{eq}}
		\label{Expression_force_term}
		\text{,}
\end{equation}
which represents the rate of change of the distribution function caused by the force\cite{Lin2017PRE}. The nine sets of kinetic moment relations can be expressed in matrix form:
\begin{equation}
		{\mathbf{C}} \cdot {\mathbf{F}} = {\mathbf{\hat F}}
		\label{Matrix_form_force}
		\text{,}
\end{equation}
where ${\mathbf F }=[F_{1} , F_{2} ,\dots,F_{N}]^{\rm{T}}$  is the column matrix of
the force term in the discrete velocity space, ${\mathbf{{\hat F}}}=[\hat F_{1},\hat F_{2},\dots,\hat F_{N}]^{\rm{T}}$ represents the column matrix of the force term in continuous space. Similarly, Eq. (\ref{Matrix_form_force}) can be expressed as:
\begin{equation}
		{\mathbf{F}} = {\mathbf{C}}^{-1} \cdot {\mathbf{{{\hat F}}}}
		\label{Expression_force_i}
		\text{,}
\end{equation}
when ${\mathbf{C}}$ is invertible. It should be pointed out that Eq. (\ref{Expression_force_i}) offers the expression of the force term.
	
Furthermore, the molar concentration $n$, mass density $\rho$, flow velocity $\mathbf{u}$, internal energy density $E$, temperature $T$, and specific heat ratio $\gamma$ are expressed by
\begin{equation}
		n = \sum\nolimits_{i}{f_{i}}
		\label{molar_concentration}
		\text{,}
\end{equation}
\begin{equation}
		\rho = m n
		\label{Density}
		\text{,}
\end{equation}
\begin{equation}
		\mathbf{u} = \dfrac{\sum\nolimits_{i}{f_{i} {\mathbf{v}}_{i }}}{n}
		\text{,}
\end{equation}
\begin{equation}
		E = \dfrac{m}{2}\sum\nolimits_{i}{f_{i} \left( |{\mathbf{v}}_{i}|^{2}+{\eta _{i}^2} \right)}
		\label{Energy}
		\text{,}
\end{equation}
\begin{equation}
		{{T}}=\dfrac{2{{E}}-{{\rho }}{{{u}}^2}}{\left( D+{{I}} \right){{n}}}
		\label{Temperature}
		\text{,}
\end{equation}
\begin{equation}
		{{\gamma }}=\dfrac{D+{{I}}+2}{D+{{I}}}
		\label{Specific_heat_ratio}
		\text{,}
\end{equation}
respectively. It is clear that the macroscopic quantities evolve as the $f_i$ are updated in the process of the discrete Boltzmann equation.
	
\subsection{Chapman-Enskog expansion}
	
The CE expansion is applied to both sides of the discrete Boltzmann Eq. (\ref{DBEquation}). The relevant variables are expanded with respect to $\varepsilon$, which corresponds to the Kn number,
\begin{equation}
		{{f}_{i}}=f_{i}^{\left( 0 \right)}+f_{i}^{\left( 1 \right)}+f_{i}^{\left( 2 \right)}+\cdots
		\text{,}
		\label{expansion_f}
\end{equation}
\begin{equation}
		\frac{\partial }{\partial t}=\frac{\partial }{\partial {{t}_{1}}}+\frac{\partial }{\partial {{t}_{2}}}+\cdots
		\text{,}
		\label{expansion_time}
\end{equation}
\begin{equation}
		\frac{\partial }{\partial r_{\alpha}}=\frac{\partial }{\partial r_{1{\alpha }}}
		\text{,}
		\label{expansion_space}
\end{equation}
\begin{equation}
		{F}_{i}={F}_{1i}
		\text{,}
		\label{expansion_force}
\end{equation}
where the nonequilibrium components $f_i^{(j)}=O(\varepsilon^j)$ and the partial derivatives $\partial/\partial {t_j}=O(\varepsilon^j)$, $\partial/\partial {r_{j\alpha}}=O(\varepsilon^j)$, for $j=1$, $2$, $\cdots$.
	
Then, by substituting Eqs. (\ref{expansion_f}) - (\ref{expansion_space}) into Eq. (\ref{DBEquation}) and equating the coefficients of the zeroth, first, second and $j$-th order terms in $\varepsilon$, we obtain
\begin{equation}
		O\left( {{\varepsilon }^{0}} \right) \text{: }  f_{i}^{\left( 0 \right)}=f_{i}^{\rm{eq}}
		\text{,}
		\label{epsilon0}
\end{equation}
\begin{equation}
		O\left( {{\varepsilon }^{1}} \right) \text{: } \frac{\partial f_{i}^{\left( 0 \right)}}{\partial {{t}_{1}}}+{{v}_{i\alpha }}\frac{\partial f_{i}^{\left( 0 \right)}}{\partial {{\alpha }_{1}}}-\frac{ma_{1\alpha}}Tf_i^{\rm{eq}}\left(v_{i\alpha}-u_\alpha\right)=-\frac{1}{\tau }f_{i}^{\left( 1 \right)}
		\text{,}
		\label{epsilon1}
\end{equation}
\begin{equation}
		O\left( {{\varepsilon }^{2}} \right) \text{: } \frac{\partial f_{i}^{\left( 0 \right)}}{\partial {{t}_{2}}}+\frac{\partial f_{i}^{\left( 1 \right)}}{\partial {{t}_{1}}}+{{v}_{i\alpha }}\frac{\partial f_{i}^{\left( 1 \right)}}{\partial {{r }_{1\alpha}}}=-\frac{1}{\tau }f_{i}^{\left( 2 \right)}
		\text{,}
		\label{epsilon2}
\end{equation}
\begin{equation}
		O\left(\varepsilon^3\right):\quad\frac{\partial f_i^{(0)}}{\partial t_3}+\frac{\partial f_i^{(1)}}{\partial t_2}+\frac{\partial f_i^{(2)}}{\partial t_1}+\frac{\partial\left(f_i^{(2)}v_{i\alpha}\right)}{\partial r_{1\alpha}}=-\frac{1}{\tau }f_{i}^{\left( 3 \right)}
		\text{,}
		\label{epsilon3}
\end{equation}
\begin{equation}
		O\left( {{\varepsilon }^{j}} \right) \text{: } \frac{\partial f_{i}^{\left( 0 \right)}}{\partial {{t}_{j}}}+\frac{\partial f_{i}^{\left( 1 \right)}}{\partial {{t}_{j-1}}}+\cdots +\frac{\partial f_{i}^{\left( j-1 \right)}}{\partial {{t}_{1}}}+{{v}_{i\alpha }}\frac{\partial f_{i}^{\left( j-1 \right)}}{\partial {{r}_{1\alpha}}}=-\frac{1}{\tau }f_{i}^{\left( j \right)}, \ j\ge 4
		\text{.}
		\label{epsilonj}
\end{equation}
	
Moreover, by utilizing the kinetic relations in Eqs. (\ref{Moment_feq0}) - (\ref{Moment_feq5,3}), and substituting $f_{i}^{\left( 0 \right)}$ with $f_{i}^{\rm{eq}}$ in Eqs. (\ref{epsilon1}) and (\ref{epsilon3}), which are computed using $\sum\nolimits_{i}{1}$, $\sum\nolimits_{i}{{{v}_{i\alpha }}}$, and
$\sum\nolimits_{i}{\left( v_{i}^{2}+\eta _{i}^{2} \right)}$, respectively. We can derive the hydrodynamic governing equations from the DBM. It can be demonstrated that in the hydrodynamic limit, the current DBM is consistent with the Euler, NS, and, Burnett equations when the discrete distribution is expanded as $f_{i}=f_{i}^{\rm{eq}}$,  $f_{i}=f_{i}^{\rm{eq}}+f_{i}^{\left( 1 \right)}$, and $f_{i}=f_{i}^{\rm{eq}}+f_{i}^{\left( 1 \right)}+f_{i}^{\left( 2 \right)}$, respectively.
	
To be specific, the Euler equations are expressed as:
\begin{equation}
		\dfrac{\partial {{\rho }}}{\partial t}+\dfrac{\partial }{\partial {r_\alpha} }\left( {{\rho }}u_{\alpha } \right)=0
		\label{EulerEquation1}
		\text{,}
\end{equation}
\begin{eqnarray}
		& \dfrac{\partial }{\partial t}\left( {{\rho }}u_{\alpha } \right)+\dfrac{\partial }{\partial r_{\beta} }\left( {{\delta }_{\alpha \beta }}{{p}}+{{\rho }}u_{\alpha }u_{\beta } \right) = \rho a_\alpha
		\label{EulerEquation2}
		\text{,}
\end{eqnarray}
\begin{eqnarray}
		& \dfrac{\partial {{E}}}{\partial t}+\dfrac{\partial }{\partial {r_\alpha} }\left( {{E}}u_{\alpha }+{{p}}u_{\alpha } \right) = \rho u_\alpha a_\alpha
		\label{EulerEquation3}
		\text{,}
\end{eqnarray}
where ${\delta_{\alpha \beta}}$ is the Kronecker function, and the pressure is
\begin{equation}
		p = n T
		\text{,}
\end{equation}
which obeys the ideal gas law.
	
Furthermore, the NS equations take the form:
\begin{equation}
		\dfrac{\partial {{\rho }}}{\partial t}+\dfrac{\partial }{\partial {r_\alpha} }\left( {{\rho }}u_{\alpha } \right)=0
		\label{NSequation1}
		\text{,}
\end{equation}
\begin{eqnarray}
		& \dfrac{\partial }{\partial t}\left( {{\rho }}u_{\alpha } \right)+\dfrac{\partial }{\partial {r_\beta} }\left( p {{\delta }_{\alpha \beta }}+{{\rho }}u_{\alpha }u_{\beta }+P_{\alpha \beta } \right) = \rho  a_\alpha
		\label{NSequation2}
		\text{,}
\end{eqnarray}
\begin{eqnarray}
		& \dfrac{\partial {{E}}}{\partial t}+\dfrac{\partial }{\partial {r_\alpha} }\left( {{E}}u_{\alpha }+{{p}}u_{\alpha }-\kappa \dfrac{\partial {{T}}}{\partial {r_\alpha} }+u_{\beta }P_{\alpha \beta } \right) = \rho u_\alpha a_\alpha
		\label{NSequation3}
		\text{,}
\end{eqnarray}
in terms of
\begin{equation}
		P_{\alpha \beta }=\mu _{\alpha \beta }\left( \dfrac{2{{\delta }_{\alpha \beta }}}{2+{{I}}}\dfrac{\partial u_{\chi }}{\partial {r_\chi} }-\dfrac{\partial u_{\alpha }}{\partial {r_\beta} }-\dfrac{\partial u_{\beta }}{\partial {r_\alpha} } \right)
		\label{P_ab}
		\text{,}
\end{equation}
with the thermal conductivity
\begin{equation}
		\kappa =\tau \dfrac{{{I}}+4}{2}\dfrac{{{\rho }}{{T}}}{{{m}}^2}
		\label{Thermal_conductivity}
		\text{,}
\end{equation}
and the dynamic viscosity
\begin{equation}
		\mu=\tau p
		\label{Dynamic_viscosity}
		\text{.}
\end{equation}
Moreover, the Burnett equations are as follows:
\begin{equation}
		\dfrac{\partial {{\rho }}}{\partial t}+\dfrac{\partial }{\partial {r_\alpha} }\left( {{\rho }}u_{\alpha } \right)=0
		\label{Burettequation1}
		\text{,}
\end{equation}
\begin{equation}
		\begin{aligned}
			\frac{\partial}{\partial t}\Big(\rho u_{\alpha}\Big) + \frac{\partial}{\partial r_\beta} \Bigg( p {{\delta }_{\alpha \beta }}+{{\rho }}u_{\alpha }u_{\beta }+P_{\alpha \beta }+H_{\alpha \beta} \Bigg)=\rho a_{\alpha}
			\label{Burettequation2}
			\text{,}
		\end{aligned}
\end{equation}
\begin{equation}
		\dfrac{\partial {{E}}}{\partial t}+\frac\partial{\partial r_\alpha}\left({{E}}u_{\alpha }+{{p}}u_{\alpha }-\kappa \dfrac{\partial {{T}}}{\partial {r_\alpha} }+u_{\beta }P_{\alpha \beta }+u_{\beta } H_{\alpha \beta}+X_{\alpha}\right)=\rho u_\alpha a_\alpha\text{,}
		\label{Burettequation3}
\end{equation}
in terms of
\begin{equation}
		\begin{aligned}
			&H_{\alpha \beta}= -\tau^2 \Bigg[2 \left( \frac{T}{m} \right)^2 \frac{\partial^2 \rho}{\partial r_{<\alpha} \partial r_{\beta>}} - \frac{2}{\rho} \left( \frac{T}{m} \right)^2 \frac{\partial \rho}{\partial r_{<\alpha}} \frac{\partial \rho}{\partial r_{\beta>}} - \frac{2\rho T}{m} \frac{\partial u_{<\alpha}}{\partial r_{\chi}} \frac{\partial u_{\beta>}}{\partial r_{\chi}} \\
			&\quad \quad\quad+ \frac{4 \rho T}{\left(2+I\right)m} \frac{\partial u_{<\alpha}}{\partial r_{\beta>}} \frac{\partial u_{\chi}}{\partial r_{\chi}} - \frac{2 \rho}{m^2} \frac{\partial T}{\partial r_{<\alpha}} \frac{\partial T}{\partial r_{\beta>}} - \frac{\rho T}{m} \frac{\partial a_{<\alpha}}{\partial r_{\beta>}}\Bigg]
			\text{,}
			\label{H_ab}
		\end{aligned}
\end{equation}
\begin{equation}
		\begin{aligned}
			&X_{\alpha} = -\tau^2 \Bigg[ \frac{4 \rho T^2}{(2+I) m^2} \frac{\partial^2 u_\beta}{\partial r_\alpha \partial r_\beta}
			- \frac{\rho T^2}{m^2} \frac{\partial^2 u_\alpha}{\partial r_\beta \partial r_\beta} - 2 \frac{\rho T}{m^2} \frac{\partial u_\beta}{\partial r_\alpha} \frac{\partial T}{\partial r_\beta}\\
			&\quad \quad\quad+ \frac{2 (6+I) \rho T}{(2+I) m^2} \frac{\partial u_\beta}{\partial r_\beta} \frac{\partial T}{\partial r_\alpha}
			- (6+I) \frac{\rho T}{m^2} \frac{\partial u_\alpha}{\partial r_\beta} \frac{\partial T}{\partial r_\beta}
			- \frac{\rho T}{m} a_\beta \frac{\partial u_{<\alpha }}{\partial r_{\beta>}} \Bigg] \text{,}
			\label{X_ab}
		\end{aligned}
\end{equation}
where the subscripts enclosed in angle brackets represents traceless tensors. Specifically,
\[
\begin{aligned}
		&\frac{\partial^2\rho}{\partial r_{<\alpha}\partial r_{\beta>}}=\frac{\partial^2\rho}{\partial r_{\alpha}\partial r_{\beta}}-\frac1{(2+I)}\frac{\partial^2\rho}{\partial r_{\chi}\partial r_{\chi}}\delta_{\alpha\beta}\text{,} \\
		&\frac{\partial u_{<\alpha}}{\partial r_{\chi}}\frac{\partial u_{\beta>}}{\partial r_{\chi}}=\frac{\partial u_{\alpha}}{\partial r_{\chi}}\frac{\partial u_{\beta}}{\partial r_{\chi}}-\frac{1}{\left(2+I\right)}\frac{\partial u_{\gamma}}{\partial r_{\chi}}\frac{\partial u_{\gamma}}{\partial r_{\chi}}\delta_{\alpha\beta}\text{,}\\
		&\frac{\partial u_{<\alpha}}{\partial r_{\beta>}}=\left(\frac{\partial u_{\alpha}}{\partial r_{\beta}}+\frac{\partial u_{\beta}}{\partial r_{\alpha}}\right)-\frac{2}{\left(2+I\right)}\frac{\partial u_{\chi}}{\partial r_{\chi}}\delta_{\alpha\beta} \text{,}\\
		&\frac{\partial a_{<\alpha}}{\partial r_{\beta>}}=\left(\frac{\partial a_{\alpha}}{\partial r_{\beta}}+\frac{\partial a_{\beta}}{\partial r_{\alpha}}\right)-\frac{2}{\left(2+I\right)}\frac{\partial a_{\chi}}{\partial r_{\chi}}\delta_{\alpha\beta} \text{,}\\
		&\frac{\partial \rho}{\partial r_{<\alpha}} \frac{\partial \rho}{\partial r_{\beta>}}=\frac{\partial \rho}{\partial r_{\alpha}}\frac{\partial \rho}{\partial r_{\beta}}-\frac{1}{\left(2+I\right)}\frac{\partial \rho}{\partial r_{\chi}}\frac{\partial \rho}{\partial r_{\chi}}\delta_{\alpha\beta}\text{,}\\
		&\frac{\partial T}{\partial r_{<\alpha}} \frac{\partial T}{\partial r_{\beta>}}=\frac{\partial T}{\partial r_{\alpha}}\frac{\partial T}{\partial r_{\beta}}-\frac{1}{\left(2+I\right)}\frac{\partial T}{\partial r_{\chi}}\frac{\partial T}{\partial r_{\chi}}\delta_{\alpha\beta}\text{.}\\
\end{aligned}
\]
	
\subsection{Nonequilibrium effects}
	
In addition to recovering the conservation equations in the continuum limit, the DBM can also capture essential nonequilibrium information that goes beyond these equations. Specifically, if the discrete equilibrium  distribution functions ${f_i^{\rm{eq}}}$ are replaced by the discrete distribution functions ${f}_i$, Eqs. (\ref{Moment_feq0}) - (\ref{Moment_feq2,0}) still hold and correspond to the conservation of mass, momentum, and energy, respectively. However, in a thermodynamic nonequilibrium state, replacing ${f_i^{\rm{eq}}}$ with ${f_i}$ may disrupt the balance in Eqs. (\ref{Moment_feq2}) - (\ref{Moment_feq5,3}). The differences between the high-order kinetic moments of ${f_i}$ and those of ${f_i^{\rm{eq}}}$ actually indicate the degree to which the physical system departs from the local equilibrium state. Mathematically, the manifestations of nonequilibrium are expressed as follows:
\begin{equation}
		\Delta_{2, \alpha \beta }=\sum\nolimits_{i}{f_{i}^{\text{neq}}{{v}_{i\alpha }}{{v}_{i\beta }}}
		\label{Delta2}
		\text{,}
\end{equation}
\begin{equation}
		\Delta_{3,1, \alpha }=
		\sum\nolimits_{i}{f_{i}^{\text{neq}}\left( v_{i}^{2}+\eta _{i}^{2} \right)v_{i\alpha }}
		\label{Delta31}
		\text{,}
\end{equation}
\begin{equation}
		\Delta_{3,\alpha \beta \chi}=
		\sum\nolimits_{i}{f_{i}^{\text{neq}}{{v}_{i\alpha }}{{v}_{i\beta }}{{v}_{i\chi }}}
		\label{Delta3}
		\text{,}
\end{equation}
\begin{equation}
		\Delta_{4,2, \alpha \beta}=
		\sum\nolimits_{i}{f_{i}^{\text{neq}}\left( v_{i}^{2}+\eta _{i}^{2} \right){{v}_{i\alpha }}{{v}_{i\beta }}}
		\label{Delta42}
		\text{,}
\end{equation}
\begin{equation}
		\Delta_{4, \alpha \beta \chi \gamma}=
		\sum\nolimits_{i}{f_{i}^{\text{neq}}{{\left( v_{i}^{2}+\eta _{i}^{2} \right)}}{{v}_{i\alpha }}v_{i\beta }v_{i\chi }v_{i\gamma }}
		\label{Delta4}
		\text{,}
\end{equation}
\begin{equation}
		\Delta_{5,3, \alpha \beta \chi }=
		\sum\nolimits_{i}{f_{i}^{\text{neq}}{{\left( v_{i}^{2}+\eta _{i}^{2} \right)}}v_{i\alpha }v_{i\beta }v_{i\chi }}
		\label{Delta53}
		\text{,}
\end{equation}
where $ f_{i}^{\text{neq}} = f_{i} - f_{i}^{\text{eq}} $ denotes the nonequilibrium part of the discrete distribution function, and the subscripts ``$m, n$" represent the contraction of an $ m $-th order tensor into an $ n $-th order tensor. Physically, $ \Delta_{2, \alpha \beta} $ represents the non-organized momentum flux, which is related to viscosity; $ \Delta_{3,1, \alpha} $ and $ \Delta_{3,\alpha \beta \chi} $ refer to the non-organized energy flux, associated with heat flux; $ \Delta_{4,2, \alpha \beta} $ corresponds to the flux of non-organized energy flux; and $ \Delta_{4,\alpha \beta \chi \gamma} $ and $ \Delta_{5,3, \alpha \beta \chi} $ reflect the nonequilibrium behaviors from the perspective of higher-order kinetic moments.

\subsection{Discretization of velocity}
	
To begin with, we should pointed out that the number of discrete velocities is not less than that of the kinetic moment relations. In other words, the least number of discrete velocities is equal to the one of kinetic moment relations, which can be achieved with the inverse matrix method. It should be mentions that the nine sets of tensor-form kinetic moment relations in Eqs. (\ref{Moment_feq0}) - (\ref{Moment_feq5,3}) and Eqs.(\ref{Moment_force0}) - (\ref{Moment_force5,3}) each contain $25$ independent components. As a result, we set $N = 25$ as the number of the discrete velocities in this work. In fact, the number of the discrete velocities is equal to that of the discrete Boltzmann equations. Consequently, the inverse matrix method provides the most efficient way for the DBM.
	
\begin{figure}
		\begin{center}
			\includegraphics[width=0.45\textwidth]{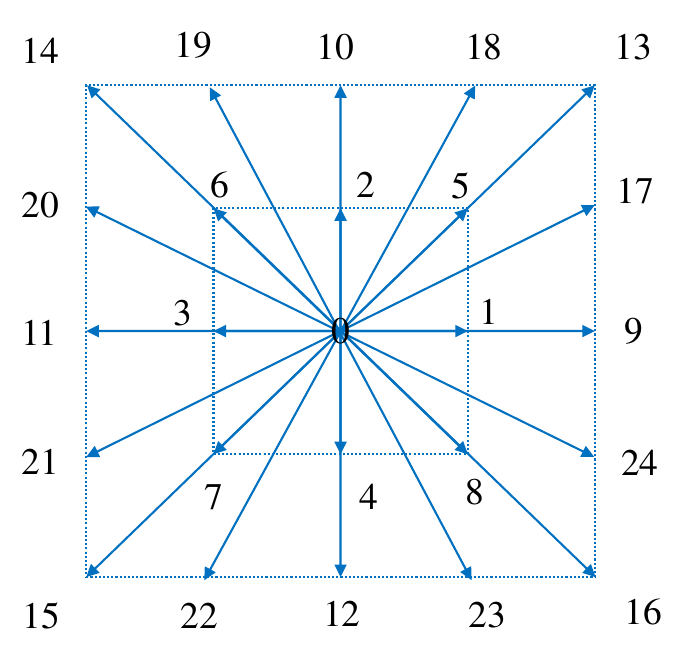}
		\end{center}
		\caption{Sketches of discrete velocities.}
		\label{Fig01}
\end{figure}
Now, let us construct the discrete velocities, D2V25, with symmetry characteristics. As delineated in Fig. \ref{Fig01}, the discrete velocities take the form,
\begin{equation}
		\mathbf{v}_{i} =
		\left\{
		\begin{array}{ll}
			0, & i = 0, \\
			\text{cyc}: v_a (\pm 1, 0), & 1 \leq i \leq 4, \\
			v_a (\pm 1, \pm 1), & 5 \leq i \leq 8, \\
			\text{cyc}: v_b (\pm 1, 0), & 9 \leq i \leq 12, \\
			v_b (\pm 1, \pm 1), & 13 \leq i \leq 16, \\
			\text{cyc}: v_b (\pm 0.5, \pm1), & 17 \leq i \leq 24 \text{,}
		\end{array}
		\right.
		\label{DVM_D2V25}
\end{equation}
where (${v}_{a}$, ${v}_{b}$) are adjustable parameters that control the magnitudes of the discrete velocities. Additionally, the parameters ${\eta}_{i}$ are assigned as ${\eta}_{0} = 0$, ${\eta}_{1 \le i \le 8}={\eta}_{a}$, ${\eta}_{9 \le i \le 16}={\eta}_{b}$, and ${\eta}_{17 \le i \le 24}={\eta}_{c}$. The values of (${ \eta }_{a}$, ${\eta }_{b}$, ${\eta }_{c}$) are also tunable.
	
It should be stressed that the values of $|\mathbf{v}_{i}|$ and ${\eta }_{i}$ can be adjusted to enhance the robustness and accuracy of the DBM. The magnitudes of ${{{\mathit{v}}}_{i}}$ can be set around the values of flow velocity $u = |\mathbf{u}|$ and sound speed $v_s = \sqrt{{\gamma} {T}/{m}}$. Furthermore, among the parameters ${\eta }_{i}$, some should be less than and the other greater than $\bar{\eta} =\sqrt{I {T} / m}$, because the extra internal energy is approximately $\dfrac{1}{2} m {\bar{\eta}}^{2}=\dfrac{1}{2} I {T}$, according to the equipartition of energy theorem.
	
To ensure efficient and accurate numerical computations while avoiding numerical oscillations and unnecessary dissipation often encountered with traditional methods, the second-order non-oscillatory and non-dissipative (NND) difference scheme \cite{Zhang1991NND} is employed for the spatial derivatives in Eq. (\ref{DBEquation}). Moreover the temporal discretization is carried out using the first-order Runge-Kutta method.
	
\section{Verification and validation}\label{sec:Verification and validation}
	
To validate the reliability of the proposed DBM, five typical benchmarks are conducted. Firstly, a free falling demonstrates the accuracy of the force term in the DBM. Secondly, a Sod shock tube shows that the DBM is applicable to compressible fluid systems. Thirdly, a sound wave indicates that the DBM can handle fluids with varying temperatures and specific heat ratios. Subsequently, thermal Couette flow confirms that the DBM is suitable for fluid systems exhibiting both hydrodynamic and thermodynamic nonequilibrium effects. Finally, the RT instability further validates the capability of the DBM in modeling complex fluid behaviors under force.

\subsection{Free falling}
First, we simulate the free fall of a fluid system to verify the correctness of the force term in this model. At the initial moment, a two-dimensional stable uniform field is given as
	$(\rho, p, T, u_x, u_y) = (1.0, 1.0, 1.0, 0, 0)$. Different accelerations $a_y$ in the $y$-direction are applied through the force term. Theoretically, the vertical flow velocity is expressed by $v_y = v_0 + a_y t$. The computational grid is set to $N_x \times N_y = 1 \times 1000$, the spatial step $\Delta x = \Delta y = 1 \times 10^{-2}$, and the time step $\Delta t = 1 \times 10^{-4}$, the relaxation time $\tau = 2 \times 10^{-4}$. Additionally, periodic boundary conditions are applied to the computational domain.
	
\begin{figure}
	\begin{center}
		\includegraphics[width=0.45\textwidth]{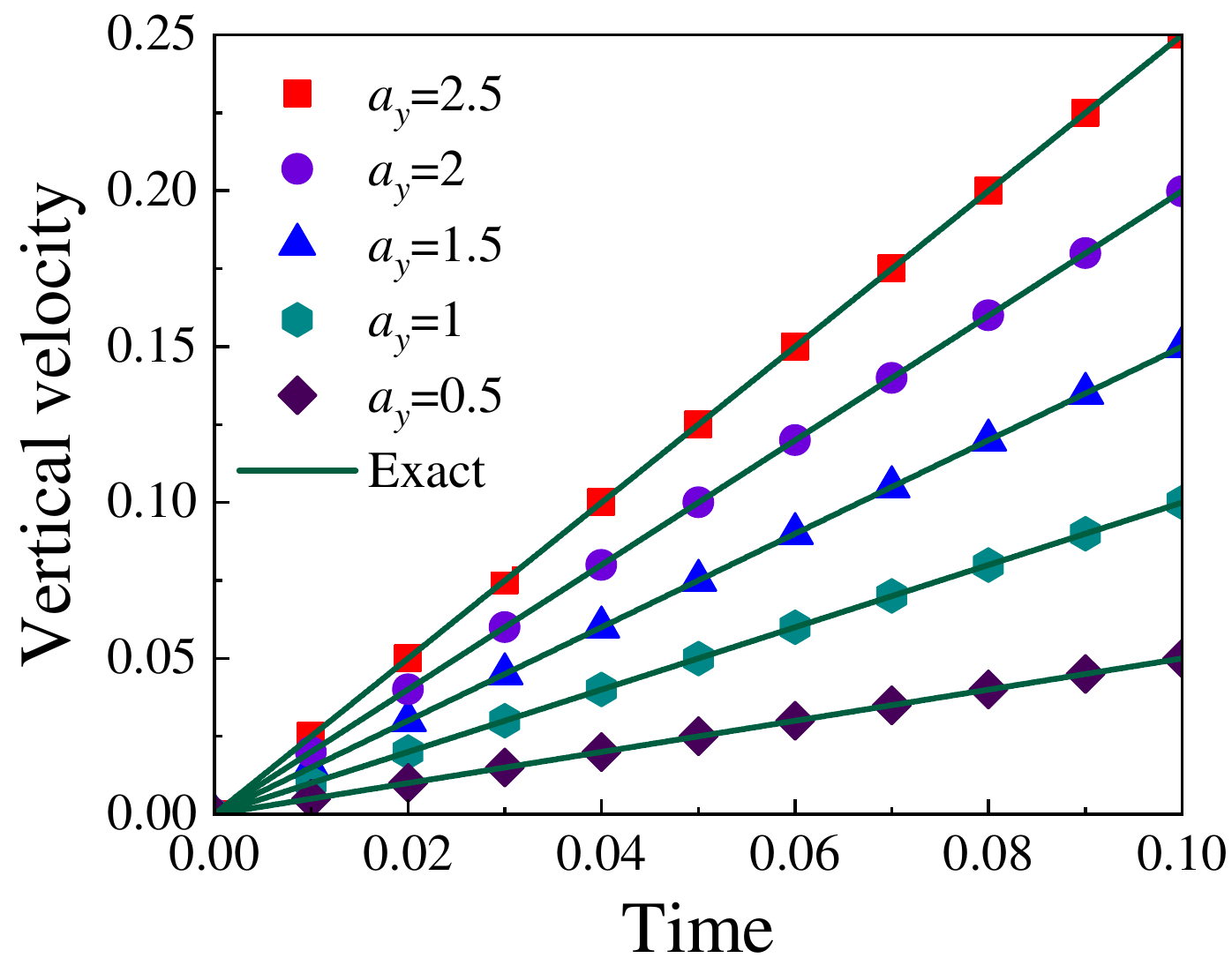}
	\end{center}
	\caption{Evolution of the vertical velocity under various accelerations during the free falling process.}
		\label{Fig02}
\end{figure}
Figure \ref{Fig02} illustrates the evolution of vertical velocity $u_y$ under various accelerations $a_y=0.5$, $1$, $1.5$, $2$, and $2.5$. The symbols represent the simulation results of the DBM, while the solid lines indicate the fitting results. The numerical results demonstrate that the velocity $v_y$ increases linearly over time, perfectly matching the theoretical predictions. The slope of each curve is proportional to the corresponding acceleration $a_y$, verifying the accuracy and correctness of the force term in the model. This confirms that the DBM can reliably simulate the hydrodynamic behavior of fluid systems under forces, providing a solid foundation for further studies on fluid systems in complex force fields.
	
\subsection{Sod shock tube}
The Sod shock tube is a fundamental benchmark for investigating shock waves phenomena, where a pressure differential is created within a closed tube, resulting in the rapid expansion of compressed gas and the formation of a shock wave. The Sod shock tube involves complex fluid dynamics, including a shock front, a rarefaction wave, and a contact discontinuity. Therefore, the Sod shock tube provides an ideal test case for evaluating the accuracy, stability, and computational efficiency of the DBM in modeling shock wave interactions and fluid dynamics under extreme conditions. To demonstrate the applicability of this DBM to high-speed compressible flows, we simulated the Sod shock tube with the following initial conditions:
\begin{equation}
		\left\{
		\begin{array}{l}
			{{\left( \rho, T, {{u}_{x}},{{u}_{y}} \right)}_{L}}=\left( 1.0,1.0,0,0 \right) 	\text{,} \\
			{{\left( \rho, T, {{u}_{x}},{{u}_{y}}\right)}_{R}}=\left( 0.125, 0.8, 0,0 \right) 	\text{,}
		\end{array}
		\right.
\end{equation}
where  $L$ and $R$ represent the initial values of the macroscopic quantities on the left ($ -1.0\leq x\leq 0.0$) and right ($ 0.0 < x\leq 1.0$) sides, respectively, away from the discontinuity interface. The grid is $N_x \times N_y = 2000 \times 2$, the spatial step $\Delta x = \Delta y = 1 \times 10^{-3}$, the time step $\Delta t = 1 \times 10^{-5}$, and the relaxation time $\tau = 7 \times 10^{-5}$. Additionally an inflow/outflow boundary condition is applied in the $x$-direction, meanwhile a periodic boundary condition is  implemented in the $y$-direction.
	
\begin{figure}
\begin{center}
\includegraphics[width=0.9\textwidth]{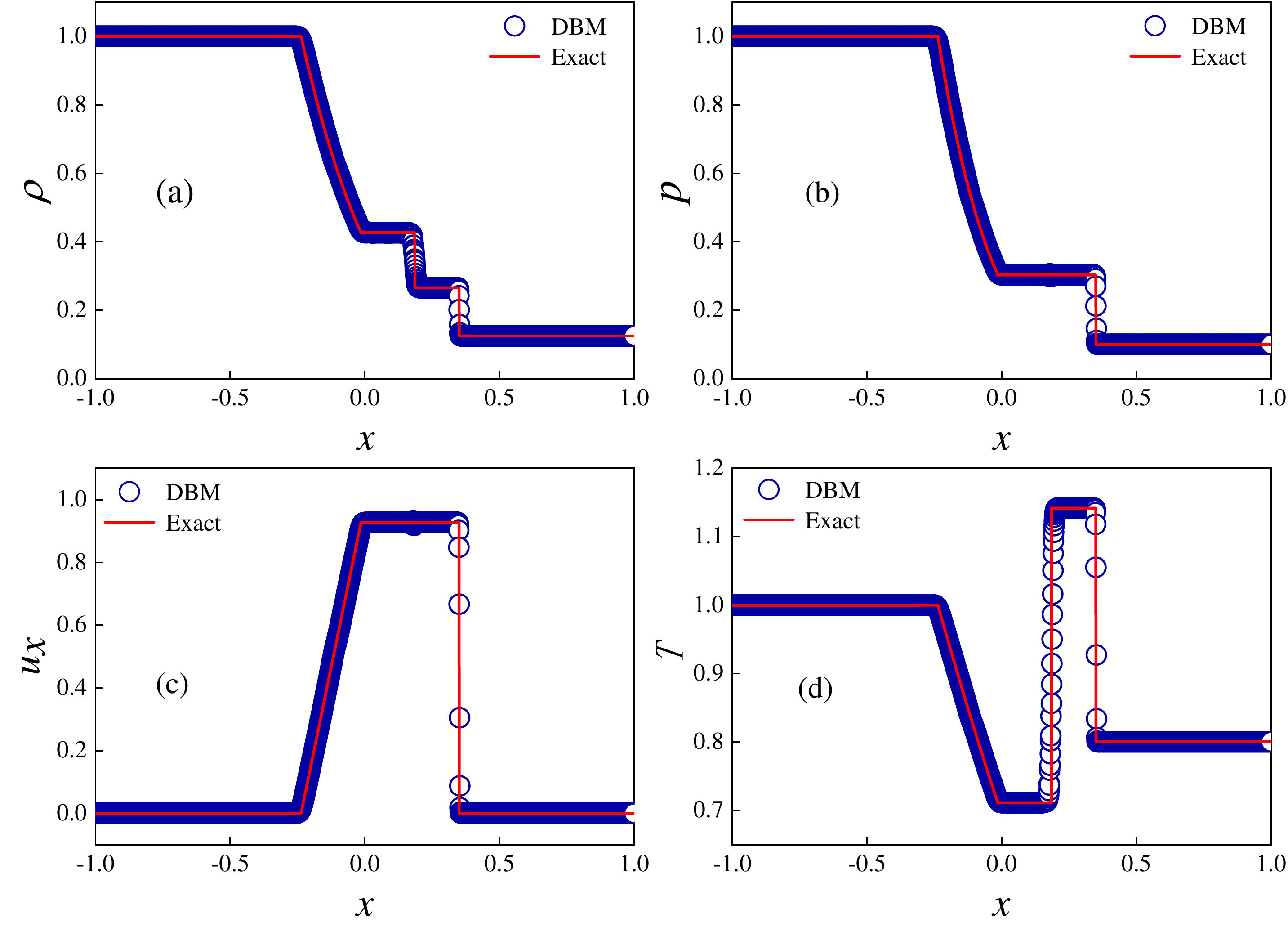}
\end{center}
\caption{Profiles of density (a), pressure (b), horizontal velocity (c), and temperature (d) at the time instant $t = 0.2$ in the Sod shock tube. }
\label{Fig03}
\end{figure}

Figures \ref{Fig03} (a) - (d) illustrate the density, pressure, horizontal velocity, and temperature along the Sod shock tube at a moment $t=0.2$. The symbols represent the simulation results obtained from the DBM, and the solid lines correspond to the Riemann analytical solutions. In the evolution of the Sod shock tube, three distinct interfaces are observed. The leftmost interface corresponds to a rarefaction wave front; the contact discontinuity separates two media with differing densities and temperatures; and the rightmost interface marks the shock front, exhibiting sharp physical gradients. The simulation results show a strong agreement with the Riemann analytical solutions, confirming that DBM effectively captures the sharp discontinuities and complex structures, such as the shock wave, contact discontinuity, and rarefaction wave, present in the Sod shock tube problem.
	
\subsection{Sound wave}
	
Now, let us prove that this model can be used to capture the sound wave. A uniform field is placed in a tube with a length $L = 2$, where the density is $\rho = 1$ and the velocity is $u = 0$. At the initial moment, a small perturbation is applied at the position $x = 0.2$ to facilitate the transmission of the sound wave. The grid is $N_x \times N_y = 2000 \times 1$, the spatial step sizes are $\Delta x = \Delta y = 1 \times 10^{-3}$,  the time step is $\Delta t = 1 \times 10^{-4}$, and the relaxation time is $\tau = 7 \times 10^{-5}$. Periodic boundary conditions are applied to the computational domain.
	
\begin{figure}
	\begin{center}
			\includegraphics[width=0.85\textwidth]{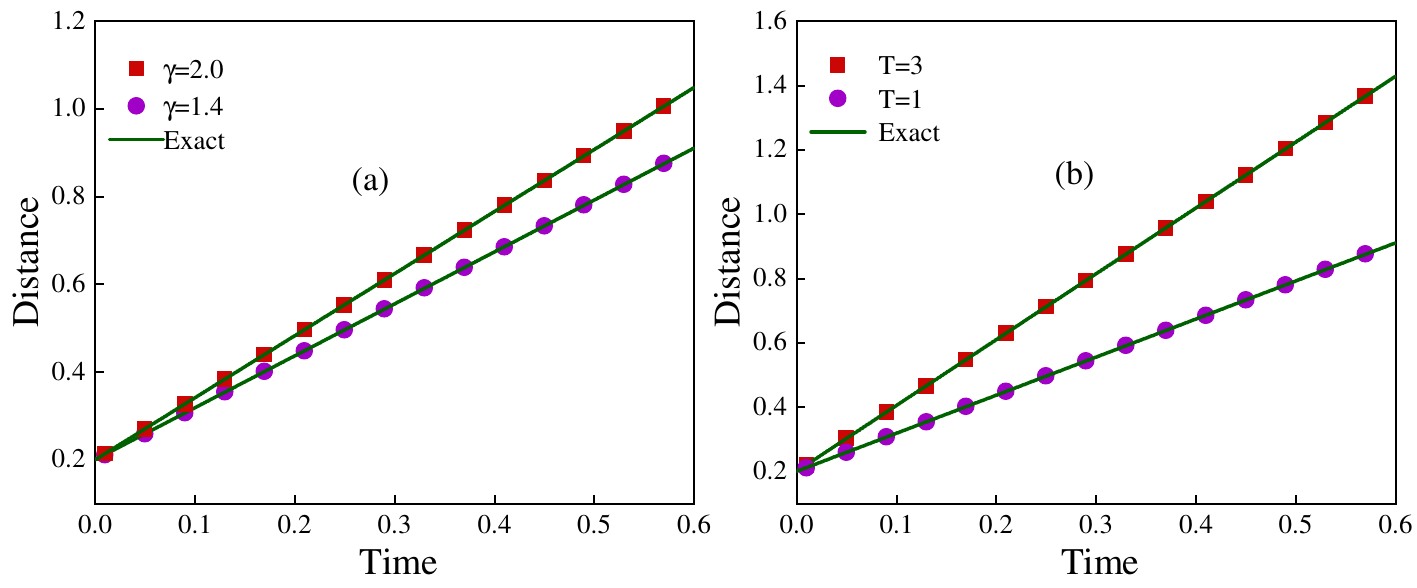}
	\end{center}
	\caption{Positions of the sound wave over time: (a) under different specific heat ratios, (b) under different temperatures.}
	\label{Fig04}
\end{figure}
	
Figure \ref{Fig04} (a) illustrates the position of the sound wave over time for two different specific heat ratios $\gamma=1.4$ and $2.0$ at a constant temperature $T=1$. In addition, Fig. \ref{Fig04} (b) shows the sound wave position for two different temperatures, $T=1$ and $3$, with a fixed specific heat ratio $\gamma=1.4$. In both figures, the simulation results align closely with the theoretical predictions, where the sound speed is given by $v_s = \sqrt{{\gamma} {T}/{m}}$. This confirms that the model accurately captures sound waves across various temperatures and specific heat ratios.

\subsection{Thermal Couette flow}
Thermal Couette flow is a classic problem in fluid mechanics, referring to the laminar motion of a fluid located between two parallel plates with different speeds. The defining characteristic of thermal Couette flow is that the fluid layers slide along the surface of the plates, with a linear velocity gradient. This makes it an ideal benchmark for validating numerical models for viscous and thermal flows. The simulation of thermal Couette flow here serves three primary purposes. The first is to confirm that the DBM is capable of handling subsonic thermal flows. The second is to illustrate its ability to capture nonequilibrium behaviors in both hydrodynamic and thermodynamic contexts. Lastly, it aims to further demonstrate that the DBM can be applicable to different specific heat ratios.
	
The initial setup is as follows: the distance between two parallel plates is $ H = 0.4 $, with the upper plate moving to the right at a constant speed $ {u}_{0} = 0.8 $, while the lower plate remains stationary. Both plates are set at a temperature $ T = 1.0 $. Initially, the fluid is at rest, with the density $ \rho = 1.0 $ and temperature $ T = 1.0 $. The computational grid is $ N_x \times N_y = 1 \times 200 $, with the spatial step $ \Delta x = \Delta y = 2.0 \times 10^{-3} $, the time step $ \Delta t = 1.0 \times 10^{-5} $, and the relaxation time $ \tau = 1.0 \times 10^{-3} $. Additionally, periodic boundary conditions are applied in the $ x $-direction, and the nonequilibrium extrapolation method is used in the $ y $-direction.
	
In theory, the exact solution for the horizontal velocity is given by:
\[u_x = \frac{y}{H} u_0 + \frac{2}{\pi} u_0 \sum_{j=1}^{\infty} \left[ \frac{(-1)^j}{j} \exp\left(-j^2 \frac{\pi^2 \mu t}{\rho H^2}\right) \sin\left( \frac{j \pi y}{H} \right) \right], \]
where $ \mu = {\tau}/{\rho T} $ represents the dynamic viscosity.
When the system reaches a steady state, the theoretical solution for the temperature in the $ y $-direction takes the form:
\[T = T_0 + \frac{u_0^2}{2c_p} \frac{y}{H} \left( 1 - \frac{y}{H} \right), \]
where $c_p = {\gamma}/{(\gamma - 1)} $ is the specific heat at constant pressure.

\begin{figure}
\begin{center}
	\includegraphics[width=0.9\textwidth]{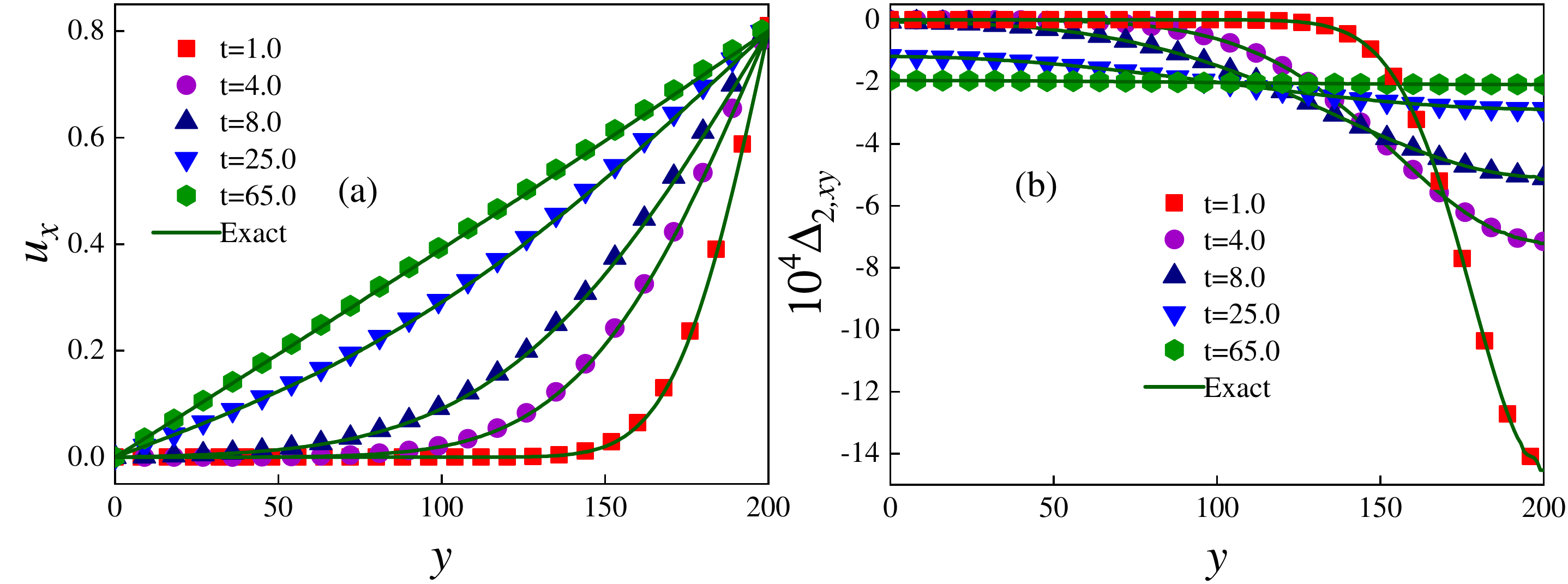}
\end{center}
\caption{Vertical distribution of $u_x$ and ${\Delta}_{2,xy}$ during the evolution of Couette flow at various time instants. The symbols represent the DBM results, and solid lines correspond to the theoretical solutions.}
\label{Fig05}
\end{figure}
\begin{figure}
\begin{center}
\includegraphics[width=0.45\textwidth]{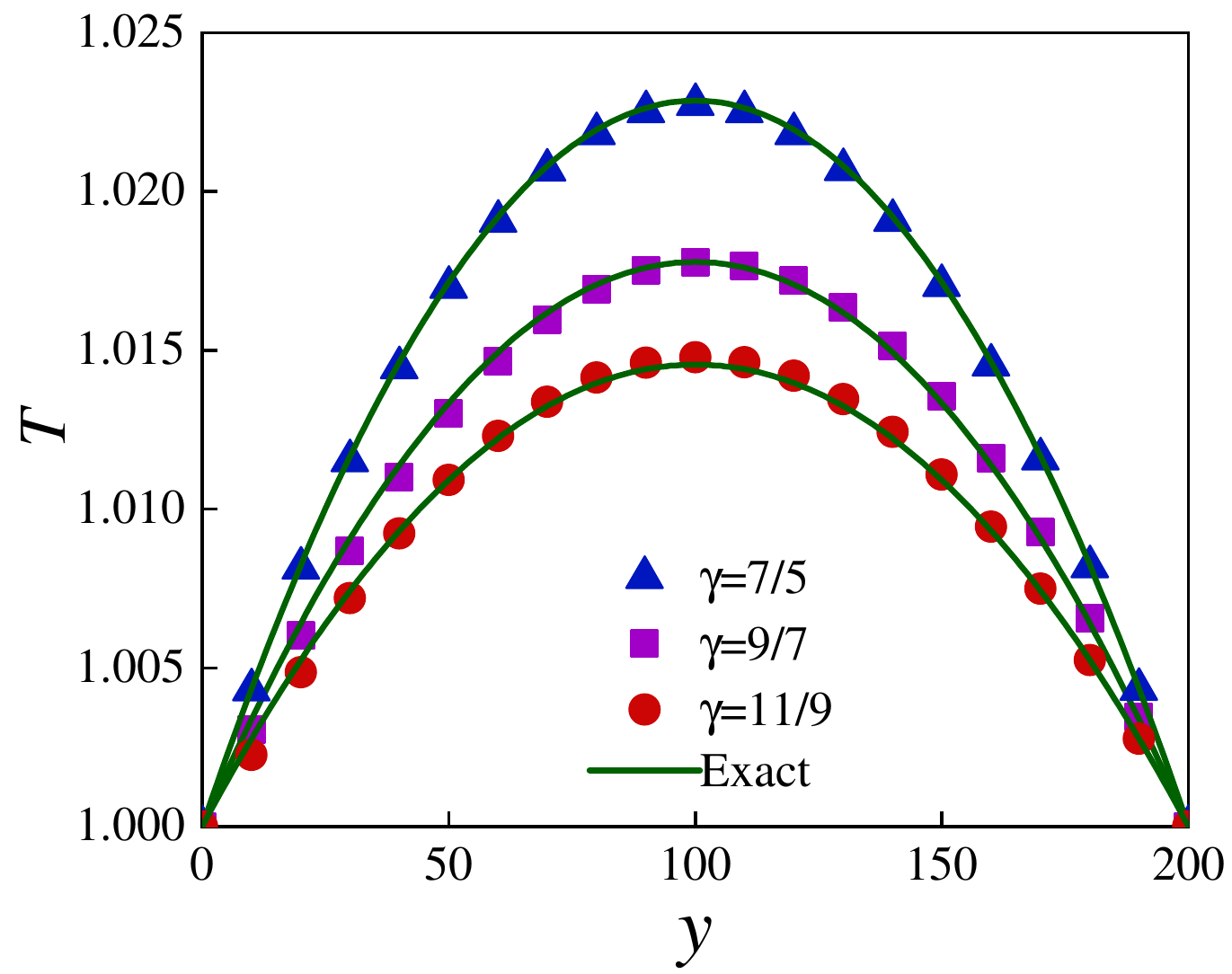}
\end{center}
\caption{Vertical distribution of the temperature in the steady Couette flow.}
\label{Fig06}
\end{figure}
Figure \ref{Fig05} presents the vertical distribution of the horizontal velocity $u_x$ (a) and the nonequilibrium quantity $\Delta_{2,xy}$ (b) in thermal Couette flow at time instants $t=1.0$, $4.0$, $8.0$, $25.0$, and $65.0$. The specific heat ratio is fixed as $\gamma=7/5$. The symbols represent the DBM results and the solid lines denote the analytical solutions. Theoretically, the analytical solution is given by ${{\Delta }_{2xy}}=- {\mu \left( {{\partial }_{y}}{{u}_{x}}+{{\partial }_{x}}{{u}_{y}} \right)}/{m}$, where $\mu$ represents dynamic viscosity \cite{lin2021multiple}. Based on the figures, we can clearly see that all simulation results align closely with the analytical solutions. Therefore, it can be concluded that the DBM effectively captures the dynamic evolution of the fluid and nonequilibrium behaviors in both hydrodynamic and thermodynamic contexts.
	
Figure \ref{Fig06} shows the temperature profiles when the thermal Couette flow achieve a steady state. The triangles, squares, and circles represent simulation results with $\gamma = 7/5$, $9/7$, and $11/9$, respectively. The solid lines represent the corresponding theoretical solutions. Clearly, the simulation results agree well with the theoretical solutions. It further shows that the DBM applies to various specific heat ratios.
	
\subsection{Rayleigh-Taylor instability}
The RT instability is a classic phenomenon in fluid mechanics that occurs at the perturbed interface between two fluids of different densities. When the heavier fluid is positioned above the lighter fluid in a gravity field, the system becomes unstable, namely, the heavier fluid sinks while the lighter fluid rises, leading to the formation of complex convective patterns. To demonstrate that the DBM can effectively simulate complex fluids under external forces, this work simulates compressible RT instability in a gravitational field and compares the results with those obtained from the D2V16 model\cite{Lai_2016PRE}.
	
The RT system is set in a two-dimensional domain $[-d, d] \times [-4d, 4d]$, with a gravitational field under constant acceleration $a = (0, -g)$. Initially, the system satisfies the hydrostatic equilibrium condition, i.e., the pressure is
	\[ \frac{\partial p(y)}{\partial y} = -g \rho(y).\]
The initial density is expressed by:
	\[\begin{aligned}
		&\quad \rho(y) = \frac{p_0}{T_u} \exp \left[ \frac{g}{T_u} \left( 2d - y \right) \right], \quad y > y_c(x), \\
		&\quad \rho(y) = \frac{p_0}{T_b} \exp \left[ \frac{g}{T_u} \left( 2d - y_c(x) \right) - \frac{g}{T_b} \left( y - y_c(x) \right) \right], \quad y < y_c(x).
	\end{aligned}\]
Here $p_0 = 1.0$ represents the initial pressure at the top of the fluid system. A cosine perturbation is applied to the interface between the two fluids: $ y_c(x) = y_0 \cos(kx) $, where $ y_0 = 0.05d $, and $ k = 2\pi/\lambda $ represents the wave number, with $ \lambda = d$ denoting the perturbation wavelength. $T_u = 1.0$ and $T_b = 4.0$ are temperatures of the upper and lower fluids far away from the interface. In addition, a tanh function is introduced to describe the transition layer, and the temperature field is set as:
\[T(y) = \frac{T_b + T_u}{2} - \frac{T_b - T_u}{2} \times \tanh \left( \frac{y - y_c(x)}{W} \right),\]
where $ W=0.02d $ denotes the width of the transition layer at the fluid interface.
Furthermore, the mesh grid is $N_x \times N_y = 256 \times 1024 $, the spatial step $ \Delta x = \Delta y = 1\times 10^{-3} $, the time step $ \Delta t = 2 \times 10^{-5} $, and the relaxation time $ \tau = 5 \times 10^{-5} $. The following boundary conditions are applied: adiabatic and no-slip boundary conditions on the top and bottom, and periodic boundary conditions on the sides.
	
\begin{figure}
\begin{center}
\includegraphics[width=1.0\textwidth]{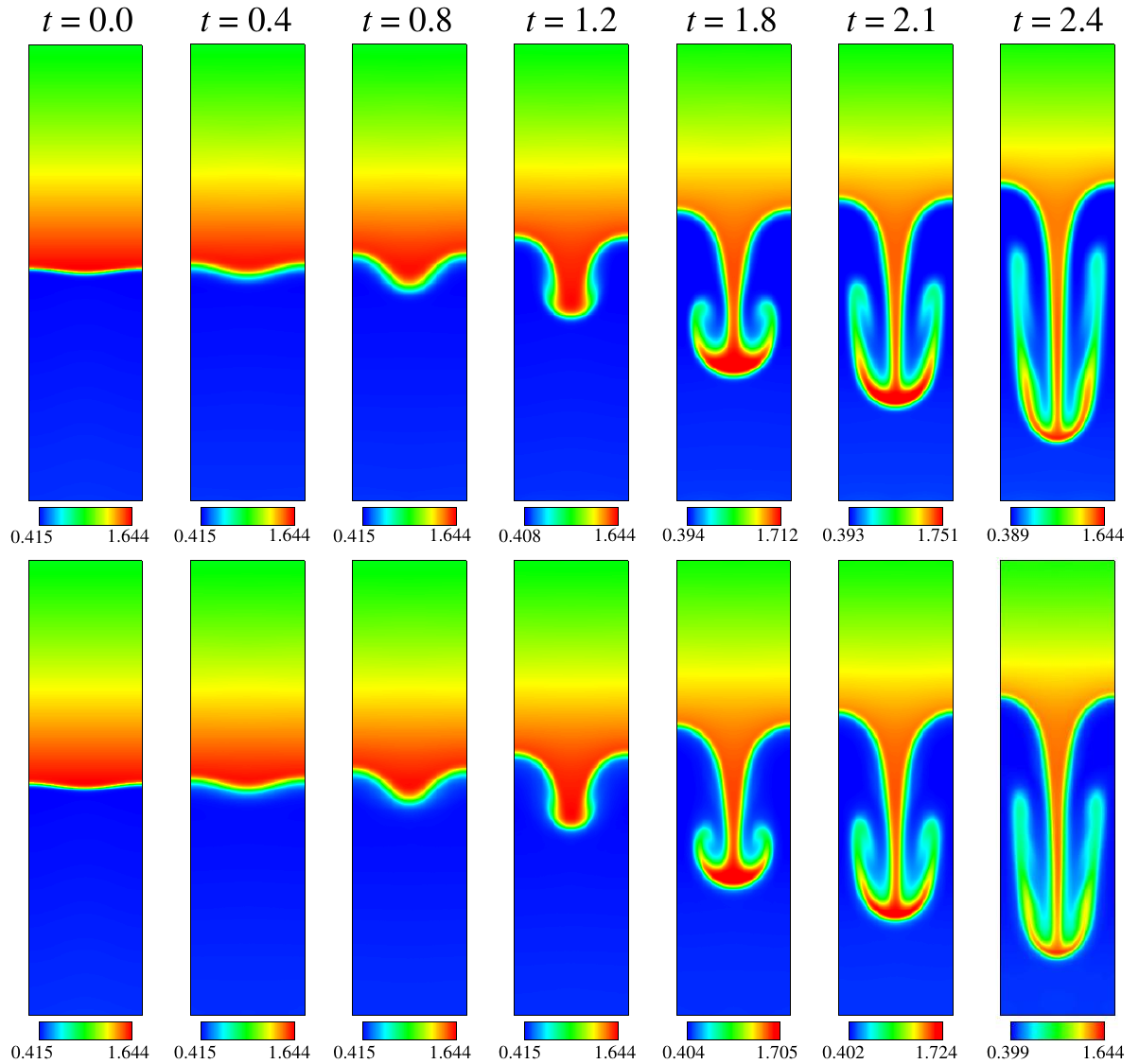}
\end{center}
\caption{Density contours in the RT process: the upper and lower rows are for D2V16 and D2V25, respectively.
		}
\label{Fig07}
\end{figure}
	
Figure \ref{Fig07}  illustrates the density contours in the evolution of the RT instability at time instants $t=0.0$, $0.4$, $0.8$, $1.2$, $1.8$, $2.1$, and $2.4$, respectively. The upper and lower rows correspond to the results obtained from D2V16 and D2V25, respectively. It can be observed that at the initial stage, the transition layer smoothens and broadens due to dissipation and diffusion effects. Before $t=0.8$, the amplitude of the initial disturbance grows exponentially while maintaining a cosine shape. As time progresses, asymmetric deformation occurs. The heavy fluid continues to move downward, forming a ``spike" shape, while the light fluid rises, forming a ``bubble" shape. In the later stages, the increase in tangential velocity difference across the interface triggers Kelvin--Helmholtz instability, causing the tail of the spike to roll up and form the characteristic ``mushroom" structure. As the fluid mixing process intensifies, the tail of the mushroom becomes increasingly blurred and elongated.
	
Both models effectively capture the key physical features of RT instability, including interface deformation, the formation of ``bubbles" and ``spikes," and the expansion of the mixing region. The similarity in their results demonstrates their accuracy in simulating compressible RT instability. Subtle variations in the density contours reveal slight differences in their handling of nonequilibrium effects. The differences stem from the fact that, in the continuum limit, the DBM with D2V16 recovers the Euler equations, while DBM with D2V25 recovers the Burnett equations. Overall, these findings affirm the reliability and robustness of the DBM in simulating complex fluid systems in force fields.
	
\section{Conclusion}\label{sec:conclusion}
	
In this paper, we present a two-dimensional Burnett-level DBM that is applicable to systems in the force fields. A discrete velocity model with $25$ velocities, designed to exhibit spatial symmetry, is constructed to ensure computational efficiency, accuracy, and stability. The CE expansion analysis demonstrates that the proposed DBM not only recovers the Burnett equations with the force term in the continuum limit, but also possesses the capability to describe diverse thermodynamic nonequilibrium behaviors. Both the equilibrium discrete distribution function and the force term satisfy $25$ independent kinetic moment relations, and their mathematical expressions can be derived via the matrix inversion method. This approach ensures high physical accuracy and computational efficiency.
	
Moreonver, the DBM is numerically validated through five benchmarks. Firstly, the free falling verifies the correctness of the force term. Secondly, the Sod shock tube confirms the model's suitability for compressible fluid systems, accurately capturing the rarefaction wave, contact discontinuity, and shock front. Thirdly, the sound wave demonstrates the model's applicability to fluids with varying temperatures and specific heat ratios. Fourthly, the thermal Couette flow ensures the model's ability to simulate subsonic thermal flows and capture nonequilibrium behaviors. Finally, the RT instability highlights the model's effectiveness in simulating complex fluids under an external force.
	
\section*{Acknowledgment}
This work is supported by National Natural Science Foundation of China (under Grant No. U2242214), Guangdong Basic and Applied Basic Research Foundation (under Grant No. 2024A1515010927), Natural Science Foundation of Fujian Province (under Grant Nos. 2021J01652, 2021J01655), Humanities and Social Science Foundation of the Ministry of Education in China (under Grant No. 24YJCZH163), and Fundamental Research Funds for the Central Universities, Sun Yat-sen University (under Grant No. 24qnpy044). This work is partly supported by the Open Research Fund of Key Laboratory of Analytical Mathematics and Applications (Fujian Normal University), Ministry of Education, P. R. China.
	
\appendix
	
\section{Matrices}\label{APPENDIXA}
	
The square matrix $\mathbf{C}$ in Eq. (\ref{Expression_feq_i}) takes the form,
\begin{equation}
		\mathbf{C}
		=
		\left(
		\begin{array}{cccc}
			{C_{11}} & C_{12} & \cdots  & C_{1N}  \\
			{C_{21}} & C_{22} & \cdots  & C_{2N}  \\
			\vdots  & \vdots  & \ddots  & \vdots   \\
			{C_{N1}} & C_{N2} & \cdots  & C_{NN}
		\end{array}
		\right)
		\text{,}
\end{equation}
with ${{C}_{1i}}=1$, $C_{2i}=v_{ix}$, $C_{3i}=v_{iy}$, $C_{4i}={v_{i}^2}+{\eta_{i}^2}$, $C_{5i}={v_{ix}^2}$,
$C_{6i}=v_{ix}v_{iy}$, ${{M}_{7i}}={v_{iy}^2}$, $C_{8i}=\left( {v_{i}^2}+{\eta _{i}^2} \right)v_{ix}$, $C_{9i}=\left( {v_{i}^2}+{\eta _{i}^2} \right)v_{iy}$,
$C_{10i}={v_{ix}^3}$, $C_{11i}={v_{ix}^2}v_{iy}$, $C_{12i}=v_{ix}{v_{iy}^2}$, $C_{13i}=v_{iy}^3$,
$C_{14i}=\left( {v_{i}^2}+{\eta _{i}^2} \right){v_{ix}^2}$, $C_{15i}=\left( {v_{i}^2}+{\eta _{i}^2} \right)v_{ix}v_{iy}$, $C_{16i}=\left({v_{i}^2} +{\eta _{i}^2} \right) {v_{iy}^2}$,  $C_{17i}={v_{ix}^3v_{iy}}$, $C_{18i}={v_{ix}^2v_{iy}^2}$, $C_{19i}={v_{ix}v_{iy}^3}$, $C_{20i}={v_{iy}^4}$, $C_{21i}={v_{iy}^4}$, $C_{22i}=\left({v_{i}^2} +{\eta _{i}^2} \right) {v_{ix}^3}$, $C_{23i}=\left({v_{i}^2} +{\eta _{i}^2} \right) {v_{ix}^2v_{iy}}$, $C_{24i}=\left({v_{i}^2} +{\eta _{i}^2} \right) {v_{ix}v_{iy}^2}$, and $C_{25i}=\left({v_{i}^2} +{\eta _{i}^2} \right) {v_{iy}^3}$. The inverse matrix $\mathbf{C}^{-1}$ can be computed using MATLAB.
\section*{References}
	
\bibliography{References}
	
\end{document}